\documentclass[10pt]{iopart}

\begin{document}
\title{Summary of Working Group 3}

\author{A. L. Kataev \dag\  and  S. Kumano\ddag}

\address{\dag\ Institute for Nuclear Research of the Academy
of Sciences of Russia, 117312, Moscow, Russia}

\address{\ddag\ Department of Physics, Saga University,
                 Saga, 840-8502, Japan}

\ead{kataev@ms2.inr.ac.ru,~kumanos@cc.saga-u.ac.jp}

\begin{abstract}
The parallel sessions of the working group 3 were devoted
to the discussions of short-baseline neutrino
physics program at neutrino factories.
First, possible studies of parton distribution
functions (PDFs), in particular nuclear and polarized PDFs,
are discussed in this summary.
Second, the extractions of $\alpha_s$ from sum rules and
structure functions of deep-inelastic neutrino-nucleon scattering,
higher-order perturbative QCD corrections and
estimates of high-twist effects are summarized.
Third, the situation of the observed NuTeV $\sin^2 \theta_W$
anomaly is discussed.
\end{abstract}




\section{Introduction}

Neutrino reactions played an important role in investigations
of hadron structure and determinations
of basic QCD and electroweak (EW) parameters.
Recently, the feasibility of constructing future neutrino factories
with intense neutrino beams was investigated in Europe, Japan and US.
In particular, proton and deuteron targets will become
available at these factories, so that the actual nucleon structure
can be investigated together with the detailed studies of
nuclear corrections.

In analyzing the neutrino-reaction processes, it is essential
to use accurate parton distribution functions (PDFs) in the nucleon.
The current status of art was summarized by Stirling \cite{Stirling}
in his plenary talk at this workshop.
It was explained that different PDF sets are constructed
from available experimental data including the data for structure
functions in unpolarized and polarized deep-inelastic scattering (DIS).
Due to the existence of various experimental data
in the wide region of $x$, the unpolarized PDFs are now well determined
from very small $x$ to relatively large $x$.
However, the situation of the polarized PDFs and nuclear PDFs is,
in particular, worse than the one of the unpolarized PDFs.
It should be stressed that these studies are really important
not only for fundamental understanding
of hadron structure but also for applications
to the reactions with heavy-ions and neutrinos.
In the first part of working group 3 (WG3) parallel sessions,
we try to understand the present situation of the nuclear
and polarized PDFs, and then possible studies at the neutrino
factories are discussed.

The physics of various neutrino-induced unpolarized processes
was also discussed at this workshop.
Because of planned huge statistics of unpolarized $\nu N$ data
with relatively low $Q^2$ at the neutrino factories,
it is becoming more realistic to investigate
non-perturbative $1/Q^2$ corrections to
Bjorken unpolarized and Gross-Llewellyn Smith sum rules,
and their comparison with available theoretical estimates.
These phenomenological investigations are closely related
to the problem of correlations between high-twist effects
and the values of QCD coupling constant $\alpha_s$, which is
extracted from experimental data in different orders of
perturbation theory. The related experiments at the neutrino factories
can put investigations of the role of twist-4 corrections
and determinations of $\alpha_s$-values
from $\nu N$ DIS characteristics on more solid ground.

More detailed studies of neutrino-induced processes can be also important
for checking the predictions of the standard EW model and independent
extractions of its parameters. A typical example is the recent
NuTeV work of Ref.~\cite{Zeller:2001hh}, which reported that
the extracted $\sin^2\theta_W$ value
from the ratios of neutral current to charged
current DIS cross-sections is in over $3\sigma$ deviation
from the result of LEP EW Working Group (LEPEWWG).
Clearly, this intriguing situation is waiting its explanation.

We summarize the presentations during the first two days of the WG3
parallel sessions.
The discussions of nuclear and polarized PDFs are reported in Sec.$\,$2.
The DIS sum rules, higher-twists, extractions of $\alpha_s$, modified
GRV98 leading-order unpolarized PDFs and the status
of NuTeV anomaly are discussed in Sec.$\,$3.
Low-energy neutrino physics was discussed at the joined sessions
with the working group 2 on the third day, and this part
is included in the summary of the working group 2 \cite{McFarland}.


\section{Parton distribution functions}

The first day of the WG3 parallel sessions
was focused on discussions of the nuclear and polarized PDFs.
Note that there are extensive reports on structure functions
and PDFs at future neutrino factories by M. L. Mangano {\it et. al.}
\cite{Mangano:2001mj,forte} and C. Albright {\it et. al.} \cite{Albright} ,
so that the reader may look at these reports for introduction.

\subsection{Nuclear PDFs}

There were  three talks on the nuclear PDFs by
J. G. Morfin, C. A. Salgado, and S. Kumano. It is known that  nuclear
effects modify the PDFs in the nucleon.
This topic has been investigated since the discovery of the EMC effect
for the structure functions $F_2$ in muon scattering.
However, due to the lack of accurate deuteron data,
nuclear effects are not seriously investigated
in neutrino-nucleus scattering. If a neutrino factory is built,
as discussed in this workshop, it could play an important role in
determining accurate nuclear PDFs. The studies are valuable for
high-energy nuclear structure physics, application to heavy-ion
phenomena, and also neutrino oscillation physics.

The nuclear PDFs have been investigated in the leading order
by many people; however, the first serious parameterization was
reported by Eskola, Kolhinen, Ruuskanen, and Salgado (EKRS) \cite{ekrs}.
In the EKRS set, the $x$ region is divided into three parts.
In the large $x$ region, the $F_2$ data fix valence-quark behavior
but sea-quark and gluon distributions are not determined.
In the medium $x$ region, the $F_2$ and Drell-Yan data constrain both
the valence and sea distributions. At small $x$,
they can determine the sea-quark distributions form the $F_2$ data,
and the valence distributions are constrained by the baryon number.
Certain $x$ dependent functional forms are assumed in these three regions,
and they are determined so as to agree with the $F_2$ and Drell-Yan (DY) data.

The PDFs in the nucleon have been obtained by $\chi^2$ analyses of various
high-energy reaction data. Unfortunately, such a technique
was not developed for the nuclear PDFs until recently. The studies
of Ref. \cite{saga01} are intended to create a simple $\chi^2$ method.
Used data set was limited to $F_2^A/F_2^D$, so that the obtained distributions
are rather different from those of the EKRS analysis.
In particular, the small-$x$ valence-quark and medium-$x$ antiquark
distributions are very different because the DY data are not included.
After the publication, the analysis has been extended
by including the DY data,
and the preliminary results show a similar $x$ dependent form to the EKRS 
except
for the gluon distribution at medium $x$. At this stage, it is not possible
to fix the gluon distribution in such an $x$ region in any case.

These nuclear PDF analyses will not be developed significantly without
new experimental data. As a new neutrino facility, the Fermilab-NuMI project
was explained by J. G. Morfin \cite{morfin}.
The unpolarized cross section for neutrino scattering is expressed
in term of three structure functions $F_1$, $F_2$, and $F_3$:
\begin{equation}
\label{cross}
\! \! \! \! \! \! \! \! \! \! \! \! \!
\frac{d^2\sigma}{dx dy} = \frac{G_F^2 s}{2\pi(1+Q^2/M_W^2)^2} \,
\bigg [ (1-y) F_2 + y^2 xF_1 \pm y (1-y/2) xF_3 \bigg ] ,
\end{equation}
where $+$ and $-$ of $\pm$ indicate neutrino and antineutrino reactions,
respectively.
The facility provides an extremely intense neutrino beam, and it
is ideal for high statistics neutrino-nucleon/nucleus experiments.
Among many physics topics at the NuMI facility,
it is a unique opportunity for investigating the nuclear PDFs, particularly,
the nuclear modification of the valence-quark distributions
through the structure function $F_3$:
\begin{equation}
\! \! \! \! \! \! \! \! \! \! \!
\frac{1}{4} \, [ \, F_3^{\nu (p+n)} + F_3^{\bar \nu (p+n)} \, ]
= u_v+d_v + (s - \bar s) + (c - \bar c)
\approx u_v+d_v
\ .
\end{equation}
It is predicted by the
parameterizations and some model studies, for example, by S. A. Kulagin
\cite{kulagin} that the valence shadowing is in general different
from the antiquark one. In the first stage of NuMI, carbon, iron,
and lead targets are installed, then the ratios $Fe/C$ and $Pb/C$
are obtained typically within a few percent
statistical errors by the three-year MINOS  run.
In the subsequent stage, $LH_2$ and $LD_2$ targets are prepared to
measure proton and deuteron structure functions. Then, the ratios
to the deuteron ($A/D$) will be obtained to find the nuclear effects.
This topic will be also investigated eventually at the neutrino
factories. There are other interesting aims of the NuMI project,
namely studies of quasi-elastics
scattering, $\sin^2 \theta_W$ issue, polarized strange-quark and charm
quark distributions \cite{morfin}.

\subsection{Polarized PDFs}

The details of the polarized PDFs are not known to the same extent
that they are understood in the unpolarized PDFs.
The reason is that  a variety of data are not yet
available. The data come from inclusive and semi-inclusive electron or muon
scattering from the polarized proton, deuteron, and $^3$He.
However, they are not enough to determine the details
of unpolarized PDFs, such as the polarized gluon distribution and
$x$-dependent shape of antiquark distributions.
Nevertheless, in the case of polarized PDFs,
the situation is better than the one for the nuclear PDF studies.
Indeed, there are
on-going polarized experiments such as RHIC and COMPASS. Moreover,
several active groups for the global $\chi^2$ analysis are working
in this area.

The results of recent global analyses of polarized data
were reported at this workshop by J. Bl\"umlein, E. Leader,
and M. Hirai. The techniques of their analysis  are almost  the same.
Bl\"umlein and B\"ottcher (BB) \cite{bb02}, Leader, Sidorov,
and Stamenov (LSS) \cite{lss02}, and Asymmetry Analysis Collaboration (AAC)
\cite{aac} have slightly different initial distributions:
\begin{eqnarray}
\! \! \! \! \! \! \! \! \! \! \!
x \Delta f_i(x, Q^2_0) & = \eta_i \, A_i \, x^{a_i} \,
            (1-x)^{b_i} \, (1 + \gamma_i \, x  + \rho_i \sqrt{x})
& \ \ \ \textstyle{\rm {in \ BB}},
\nonumber \\
\! \! \! \! \! \! \! \! \! \! \!
\Delta f_i(x, Q^2_0) & = \eta_i \, A_i \, x^{a_i} \, f_i(x, Q^2_0)
& \ \ \ \textstyle{\rm {in \ LSS}},
\label{eqn:initial}
\\
\! \! \! \! \! \! \! \! \! \! \!
\Delta f_i(x, Q^2_0) & = A_i \, x^{\alpha_i} \,
                (1 + \gamma_i \, x^{\lambda_i} ) \, f_i(x, Q^2_0)
& \ \ \ \textstyle{\rm {in \ AAC}}.
\nonumber
\end{eqnarray}
Here, $f_i$ is the corresponding unpolarized distribution in the nucleon.
There are also  other recent studies on the polarized PDFs \cite{recent0102}.
These distributions are evolved to the experimental $Q^2$ points of
the spin asymmetry $A_1$ or $g_1$. Then, the parameters
in Eq. (\ref{eqn:initial}) are determined so as
to minimize the $\chi^2$ value.
In the updated analysis of the LSS, the SLAC-E155 data are included
by taking care of the positivity constraint and using updated
values for the axial charges $a_3$ and $a_8$.
It is important that the BB analysis demonstrated  error bands for
the polarized PDFs, so that the uncertainties of the present
distributions become clear. In particular, the error band is very
large for the polarized gluon distribution, which should be determined
more precisely by future measurements.
An error analysis  was also made by AAC. The AAC error bands show similar
tendency; however, the details are different due to the different
choice of the functional form.
For example, the gluon error is larger for AAC than the one for BB.

Possibilities of determining the polarized strange-quark distribution
in neutrino reactions were discussed by K. Sudoh, and
charmed meson detectors
were  discussed by P. J. M. Soler.
Semi-inclusive $D$ and $\bar D$ production cross sections were investigated
in polarized neutrino-proton reactions \cite{sudoh}. Four different
parameterizations, AAC, BB, GRSV, and LSS, are used for calculating
the polarization asymmetry. Because a charm quark is produced partially
from a strange quark in the nucleon, the studies indicate that the polarized
strange-quark distribution could be measured by these reactions.
Experimentally, silicon detectors have been developed by the NOMAD-STAR
group \cite{soler} for the charm identification in neutrino reactions,
and it was installed inside the NOMAD experiment.
 From about 11,500 $\nu_\mu$ charged-current events, 45 charmed mesons were
obtained. The studies indicate that the silicon detectors could be
used at the future neutrino factories for identifying the charmed mesons,
so that they could contribute significantly to the PDF studies.

The situation of the HERMES experiment and possible
neutrino experiments were presented by T.-A. Shibata.
Near-future polarized experiments were discussed by N. Saito.
The HERMES experiments measured semi-inclusive hadron production
cross sections in polarized electron-nucleon scattering,
and the data enabled them to investigate flavor decomposition
of the polarized PDFs \cite{shibata}. Furthermore, the single
spin azimuthal asymmetry measurements indicated chiral-odd distributions.
 From his experience of these experiments, he suggested that
a future neutrino factory should have detectors with good particle
identification in the final state for precise determination
of the polarized PDFs.
Next, N. Saito explained the present and near-future experiments to
study the polarized PDFs \cite{saito}. The present data are mainly taken
for the inclusive structure function $g_1$ and some semi-inclusive
reactions, so that they provide us a limited information
for the gluon polarization. However, since
RHIC, HERMES, COMPASS, and TESLA-N projects intend to measure
$\Delta g(x)/g(x)$ accurately, the situation should become clear
in the near future.
In particular, he demonstrated that the RHIC $\Delta g$
measurements provide a strong constraint for the gluon polarization.
Indeed, after including virtual RHIC data in the global fit,
the $\Delta g$ error becomes significantly smaller.
In addition, the $W$ production measurements
should provide a constraint for the polarized antiquark distributions.

As for the polarized PDFs and structure functions at the neutrino factories,
there were discussions by E. Leader and G. Ridolfi.
There are extensive studies of polarized structure functions and polarized
PDFs, which could be obtained from
the possible European neutrino factory \cite{Mangano:2001mj,forte}.
In addition to the polarized structure functions
$g_1$ and $g_2$ in electron or muon scattering, there are new
ones, namely  $g_3$, $g_4$, and $g_5$, in polarized neutrino
scattering. In should be noted, however, that
there are different definitions for these functions among researchers,
and they are summarized in Ref. \cite{g345}.
In the convention of Refs. \cite{Mangano:2001mj,forte,g345}, $g_4$ and $g_5$
are related by $g_4=2x g_5$ in the leading order.
Neglecting the $g_2$ and $g_3$ terms by taking the high-energy limit,
we have \cite{Mangano:2001mj,forte}
\begin{equation}
\! \! \! \! \! \! \! \! \! \! \! \! \! \! \! \! \! \! \! \! \! \! \!
\frac{d^2\Delta\sigma^{\lambda_\ell}}{dx \, dy}
= {G^2_F  \over \pi (1+Q^2/m_W^2)^2}{Q^2\over xy}
   [ -\lambda_\ell\, y (2-y)  x g_1- (1-y) g_4- y^2 x g_5 ] ,
\end{equation}
where $\lambda_\ell$ is the lepton helicity, and
$\Delta\sigma$ is the difference between the polarized
cross sections: $\Delta\sigma= \sigma_{\lambda_p=-1}-\sigma_{\lambda_p=+1}$
with the proton helicity $\lambda_p$.

The charged-current structure functions $g_1$ and $g_5$ are expressed in term
of the polarized PDFs in the leading order:
\begin{eqnarray}
\! \! \! \! \! \! \! \! \! \! \! \! \!
& g_1^{W^+} =  \Delta\bar u + \Delta d + \Delta s + \Delta\bar c,
\ \ \ \
& g_1^{W^-} = \Delta u +\Delta\bar d + \Delta\bar s + \Delta c,
\nonumber \\
\! \! \! \! \! \! \! \! \! \! \! \! \!
& g_5^{W^+} = \Delta\bar u - \Delta d -\Delta s +\Delta \bar c,
\ \ \ \
& g_5^{W^-} = -\Delta u + \Delta \bar d + \Delta \bar s - \Delta c .
\end{eqnarray}
A combination of the $g_1$ structure functions becomes
the flavor singlet distribution:
\begin{eqnarray}
\! \! \! \! \! \! \! \! \! \! \! \! \! \! \! \! \! \! \! \! \! \!
& \Delta\Sigma (x) & = (g_1^{W^+ +W^-})_p=(g_1^{W^+ +W^-})_n
\nonumber \\
& & =\Delta u+\Delta\bar u +\Delta d+\Delta\bar d
+\Delta s+\Delta\bar s +\Delta c+\Delta\bar c ,
\label{eqn:ds}
\end{eqnarray}
where isospin symmetry is used for the distributions in the neutron.
It is especially important that the spin content
$\Delta\Sigma = \int dx \Delta\Sigma (x)$ is found directly.
In the present situation, the spin content is determined
by the $\chi^2$ analysis of electron and muon scattering data
together with the first moments $\int dx \Delta u_v$ and
$\int dx \Delta d_v$, which are fixed by semi-leptonic decay data.
According to Eq. (\ref{eqn:ds}), we do not have to rely on such
low-energy data.
As noticed in Ref. \cite{aac}, the accurate determination
of $\Delta \Sigma$ is not possible at this stage because
the polarized antiquark distribution $\Delta \bar q(x)$ cannot
be determined at small $x$ from the present data.
The neutrino reactions should provide valuable information
on the spin content. Combining these structure functions,
we obtain, for example,
\begin{eqnarray}
\! \! \! \! \! \! \! \! \! \! \! \! \! \! \! \! \! \! \! \! \! \!
\! \! \! \! \! \! \! \! \! \! \! \! \! \!
& (g_5^{W^+ +W^-})_p=(g_5^{W^+ +W^-})_n
  = -\Delta u+\Delta\bar u -\Delta d + \Delta\bar d
-\Delta s+\Delta\bar s -\Delta c+\Delta\bar c,
\nonumber\\
\! \! \! \! \! \! \! \! \! \! \! \! \! \! \! \! \! \! \! \! \! \!
\! \! \! \! \! \! \! \! \! \! \! \! \! \!
& (g_5^{W^+ -W^-})_p+(g_5^{W^+ -W^-})_n
= -2 \, (\Delta s+\Delta\bar s) + 2 \, (\Delta c + \Delta\bar c) ,
\\
\! \! \! \! \! \! \! \! \! \! \! \! \! \! \! \! \! \! \! \! \! \!
\! \! \! \! \! \! \! \! \! \! \! \! \! \!
& (g_1^{W^+ -W^-})_p+(g_1^{W^+ -W^-})_n
  = 2 \, (\Delta s-\Delta\bar s) - 2 \, (\Delta c - \Delta\bar c) .
\nonumber
\end{eqnarray}
Therefore, the first $g_5$ combination is especially useful for determining
the polarized valence-quark distributions. Furthermore, the second $g_5$
combination indicates the polarized strange- and charm-quark distributions.
If the $g_1$ data are accurate enough, it could be possible to
find the difference between $\Delta s$ and $\Delta \bar s$.
The feasibility of measuring these structure functions $g_1$ and $g_5$
was investigated for the European neutrino factory in Ref. \cite{forte},
and possible errors are estimated in the region $x>0.1$.
In particular, G. Ridolfi showed in his talk
that the first moments of C-even distributions
($\Delta q+\Delta \bar q$) could be improved by an order of magnitude
in comparison with the present uncertainties, whereas those of C-odd 
distributions
($\Delta u-\Delta \bar u$, $\Delta d-\Delta \bar d$) are determined
at the level of a few percent.
In this way, measurements of $g_1$ and $g_5$ should provide
important information for the polarized PDFs.


\section{Unpolarized deep-inelastic scattering}
The second part of the WG3 sessions was devoted to
the discussions of the physical information, which can be obtained
from theoretical and experimental considerations of characteristics of
unpolarized DIS. These characteristics are
related to the differential cross-sections in Eq. (\ref{cross})
where $0\leq x\leq 1$, $y=E_{had}/E_{\nu}$, and
$0\leq y\leq 1/(1+\frac{xM_W}{2E_{\nu}})$.
Because of the large variation of $y$ at the neutrino factories
\cite{Mangano:2001mj}, it is possible to extract
not only the standard structure functions (SFs) $F_2^{\nu N}$ and
$xF_3^{\nu N}$ but also $F_1^{\nu N}$ from the cross-sections
of Eq. (\ref{cross}). This procedure might allow us to push ahead new
QCD studies, which aim at a direct analysis of
scaling violation in $F_1^{\nu N}$ SF data.
Moreover, in addition to two well-known sum rules of $\nu N$ DIS,
namely to the Adler sum rule
\begin{equation}
\label{A}
\! \! \! \! \! \! \! \! \! \! \! \! \! \! \! \! \! \!
I_{F_2}=
\int_0^1 \frac{dx}{x}\bigg[F_2^{\nu n}(x,Q^2)-
F_2^{\nu p}(x,Q^2)\bigg]= 2
\end{equation}
and the Gross--Llewellyn Smith sum rule
\begin{equation}
\label{GLS}
\! \! \! \! \! \! \! \! \! \! \! \! \! \! \! \! \! \!
I_{F_3}=
\frac{1}{2}\int_0^1\bigg[F_3^{\nu n}(x,Q^2)
+F_3^{\nu p}(x,Q^2)\bigg] = 1 -\frac{\alpha_s}{\pi}-...+
O\bigg(\frac{1}{Q^2}\bigg) \ ,
\end{equation}
it will also be   possible to verify
the Bjorken unpolarized sum rule
\begin{equation}
\label{Bju}
\! \! \! \! \! \! \! \! \! \! \! \! \! \! \! \! \! \!
I_{F_1}=
\int_0^{1}dx\bigg[F_1^{\nu n}(x,Q^2)-F_1^{\nu p}(x,Q^2)\bigg]=1-
\frac{2}{3}\frac{\alpha_s}{\pi}-....+O\bigg(\frac{1}{Q^2}\bigg) ~.
\end{equation}
Note that,  contrary to the exact theoretical
expression for Eq. (\ref{A}), the integrals $I_{F_3}$ and $I_{F_1}$
contain
perturbative QCD corrections, calculated explicitly in
the $\overline{MS}$-scheme up to order
$\alpha_s^3$ in Refs.\cite{Larin:tj} and \cite{Larin:1990zw}
respectively and estimated by different ways  at  the $\alpha_s^4$
level (see Refs.\cite{Kataev:1995vh,Samuel:1995jc,Broadhurst:2002bi}).
Therefore, experimental measurements  of these DIS sum rules can
provide useful information on the value of the QCD coupling constant
$\alpha_s$. Moreover, while the   existing  sets  of data for
the  $xF_3$ SF, including the most precise published
one provided by the  CCFR collaboration
\cite{Seligman:mc}, were already used for the extraction
of $I_{F_3}$  at different
$Q^2$ bins \cite{Kim:1998ki} (see Ref. \cite{Kataev:1994rj}
for the discussion of the $Q^2$ behavior
of $I_{F_3}$, extracted from previous CCFR data),
new experiments at the neutrino factories can give the possibility of
the first measurement of the  $I_{F_1}$ sum rule.
It is interesting from the point of view of an independent determination
of $\alpha_s$  (see Ref. \cite{Mangano:2001mj}). This problem was
discussed at this workshop in the talk of Ref. \cite{Alekhin:2002pj}.

It should be stressed that theoretical $\alpha_s$ ambiguities,
which result from the comparison of the QCD predictions
with DIS experimental data,
and with the sum rules $I_{F_3}$ and $I_{F_1}$ in particular,
are related to the uncertainties in the
non-perturbative twist-4 corrections. They  manifest themselves as
the $O(1/Q^2)$-contributions to the corresponding theoretical
predictions. In the case of the mentioned  sum rules, the  twist-4
model-independent
matrix elements are known \cite{Shuryak:1981kj}.
The problem of getting concrete
numbers for these matrix elements was discussed in detail in the
talk of Weiss \cite{Weiss}. He presented the preliminary
calculations of twist-4 contributions to $I_{F_3}$
and $I_{F_1}$ within the framework
of the  instanton-vacuum model, developed
in Ref. \cite{Diakonov:1983hh}.
In the case of $I_{F_3}$, the reported estimates
turned out
to be in good agreement with the ones obtained in Ref.\cite{Braun:1986ty}
with the help of three-point function QCD sum rules.
Within
existing theoretical uncertainties, this result of
Ref. \cite{Braun:1986ty} is supported by the independent
three-point function QCD sum rule analysis of   Ref. \cite{Ross:1993gb}.
In the case of another twist-4 operator, which contributes
to $I_{F_1}$, the estimate reported by Weiss  also confirmed
the results of Ref. \cite{Braun:1986ty} obtained with the help of
three-point function QCD sum rules.
As demonstrated in Ref. \cite{Alekhin:2002pj}, the errors
of the twist-4 contribution to  $I_{F_1}$
are playing the dominant role  in the uncertainties of
$\alpha_s(M_Z)$, extracted from the  $Q^2$-behavior of
this sum rule with the next-to-leading order (NLO) PDF set of
Ref. \cite{Alekhin:2000ch}. Indeed, in the process of normalizing the
theoretical
expression for $I_{F_1}$ to   $Q^2=4$ GeV$^2$
and $Q^2=10$ GeV$^2$, the authors of Ref. \cite{Alekhin:2002pj} obtained
$\Delta^{HT}\alpha_s(M_Z)=0.012$ and $\Delta^{HT}\alpha_s(M_Z)=0.007$.
It is now important to fix in more detail the  theoretical errors
in the  calculations of the twist-4 contributions to $I_{F_1}$,
performed by different methods in Refs. \cite{Braun:1986ty}
and \cite{Weiss}.

However, the inclusion of higher-order perturbative QCD corrections
into the $\alpha_s(M_Z)$ determination
procedures can result in the appearance of a problem
of correlations between perturbative and non-perturbative QCD  contributions.
This property was discovered in the process of the next-to-next-to-leading
order (NNLO) fits to $xF_3$ data of the CCFR collaboration \cite{Kataev:1997nc}
with the NNLO corrections to anomalous dimensions of even non-singlet 
moments for
$F_2$ SF calculated in Ref. \cite{Larin:1996wd}.
(The summary of more detailed recent NNLO fits to $xF_3$ data 
\cite{Kataev:2001kk}
  was reported to WG3  in the talk of Ref. \cite{Kataev:2002wr}. 
The fits performed in  Ref.\cite{Kataev:2001kk} are based on the application
  of the  NNLO corrections to anomalous dimensions of odd moments
  for $xF_3$ SF calculated in Ref. \cite{Retey:2000nq}.)
A similar pattern, discussed in the contribution of Ref.\cite{Bodek},
was independently revealed in Ref. \cite{Yang:1999xg}
in the process of the NNLO fits to SLAC, NMC and BCDMS data
for the $F_2$ SF. As was argued further on \cite{Contreras:2002kf}
(see the talk by Cvetic \cite{Cvetic}),
the application of the Borel resummation technique, together with
the incorporation of the the $1/Q^2$-term typical to the infrared renormalon
approach (for a review, see Ref. \cite{Beneke:1998ui}),
leads to a minimization of the role of this power correction
in the procedure of extracting $\alpha_s(M_Z)$ from the data for $I_{F_3}$
at $Q^2\leq 4$ GeV$^2$.
The result in Ref. \cite{Cvetic},
\begin{equation}
\label{resummed}
\alpha_s(M_Z)=0.1167^{+0.0128}_{-0.0118}~(\rm{exp}) \pm 0.0008~ (\rm{th})~,
\end{equation}
which follows from the resumed perturbative expression for $I_{F_3}$,
turned out to be closer to the world average value of
$\alpha_s(M_Z)=0.1184\pm 0.0031$
\cite{Bethke:2000ai} than its analogue:
\begin{equation}
\label{truncated}
\alpha_s(M_Z)=0.114^{+0.005}_{-0.006}~({\rm stat})^{+0.007}_{-0.009}~
(\rm{syst})\pm 0.005~(\rm{th})
\end{equation}
obtained in Ref. \cite{Kim:1998ki}  by the  CCFR/NuTeV collaboration
from the truncated NNLO
perturbative series for $I_{F_3}$, supplemented with
the estimate of  twist-4 power correction from
Ref. \cite{Braun:1986ty}.
Note that from  our point of view, the theoretical uncertainties in
Eq. (\ref{resummed}) might  be underestimated. Moreover, in
view of different opinions on the role of non-perturbative effects in the
extraction of $\alpha_s$ from the $I_{F_3}$ sum rule
and of the possible appearance of more precise data for $xF_3$
at the neutrino factories, it would be useful to clarify
the role of twist-4 terms in the $\alpha_s(M_Z)$ determination
from the $I_{F_3}$ sum rule.

Let us return to the problem of parameterizing non-perturbative effects
to DIS  SFs.
Contrary to the discussed sum rules, the complete form of
dynamical twist-4 corrections in both $xF_3$ and $F_2$
is unknown. Therefore,  various  models for $1/Q^2$
contributions  are generally used. The most popular one was constructed
in Ref. \cite{Dasgupta:1996hh} with the help of the infrared
renormalon (IRR) approach. Within this approach,
the $1/Q^2$ contributions to SFs are defined as
$F_i^{HT}/Q^2=(A_2^{'}/Q^2)\int_x^1(dz/z)C_i(z)q(x/z,Q^2)$, where
$q=u_v+d_v$ indicates
the valence contributions, $C_i(z)$ is a part of the coefficient function
which is calculated from the chain of quark, gluon and ghost 
loop insertions into the internal gluon line of the corresponding
Born diagrams, and  $A_2^{'}$
is the free parameter which should be extracted from the fits to
experimental data.

Here we recall that in the era when experimental results
for $xF_3$ were not precise enough, the
concrete fits of Ref. \cite{Abbott:1979as} were not able to make
an unambiguous choice between perturbative and non-perturbative
sources of scaling violation. At present, thanks to the  increase in
precision of $\nu N$ DIS experiments, it has become
possible to separate the two mechanisms
of scaling violation in the process of NLO fits. Indeed, the most
recent NLO analysis of CCFR data for $xF_3$ SF \cite{Kataev:2001kk}
described in the talk of Ref. \cite{Kataev:2002wr}
allowed the extraction of both $\alpha_s(M_Z)$ of the $\overline{MS}$-scheme
and the free parameter of the IRR model $A_2^{'}$ with reasonable
error bars:
\begin{eqnarray}
\label{results}
\alpha_s(M_Z)&=&0.120 \pm 0.002~(\rm{stat})\pm 0.005~ (\rm{syst})
\\ \nonumber
&&\pm 0.002~(\rm{thresh})^{+0.010}_{-0.006}~ (\rm{scale}) \\
\label{IRR}
A_2^{'}&=&-0.125 \pm 0.053~{\rm GeV}^2
\ .
\end{eqnarray}
The theoretical errors in $\alpha_s(M_Z)$ reflect the uncertainties in
passing the threshold of production of $b$-quarks and the ambiguities in
choosing renormalization and factorization scales. It is interesting that
the NLO fits to $F_2$ data, performed in Ref. \cite{Yang:1999xg} with
the help of the MRS(R2) PDF set \cite{Martin:1996as} result in the value
of $A_2^{'}=-0.104\pm0.005$ GeV$^2$, which is in agreement
with the result of Eq. (\ref{IRR}).

As shown in the talks of Refs. \cite{Kataev:2002wr,Bodek}, the incorporation
of the NNLO corrections into the fits to $xF_3$ and $F_2$ data makes
the detection of dynamical $1/Q^2$- contributions more problematic,
provided that the NNLO values of $\alpha_s(M_Z)$ are lying not far from
the world average result of $\alpha_s(M_Z)\approx 0.118$.
Indeed, the IRR model parameter obtained from the NNLO fits
to CCFR'97 $xF_3$ data is comparable with zero, within the statistical
error bars:  $A_2^{'}=-0.013\pm0.051$ GeV$^2$
(see e.g. Ref. \cite{Kataev:2001kk}). 
A similar value, namely $A_2^{'}=-0.0065\pm 0.0059$ GeV$^2$, 
was obtained from the NNLO fits to $F_2$ SF data \cite{Yang:1999xg}
based on application of the recent estimates of the NNLO DGLAP splitting 
functions for the NNLO PDFs \cite{vanNeerven:1999ca}.
Moreover, as shown in
Refs. \cite{Santiago:2001mh,Kataev:2001kk,Maxwell:2002mt}, the effect of
minimization of the
$1/Q^2$ contributions to $xF_3$ SF extracted at the NNLO does not depend
on  the model chosen for their parameterization. Thus,
we are  returning  to the situation of 1979 at a new level
of understanding. Indeed,  we can conclude that, in order to detect
at the NNLO signals from dynamical $1/Q^2$-contributions to
the DIS structure functions,
we need more precise experimental data, which can be obtained in
the future at the neutrino factories.

It is planned that the neutrino factories
will cover the region with low $Q^2$. In the second part of the
contribution of Ref.\cite{Bodek}, based on the second talk of Bodek,
the modification of GRV98 LO PDFs \cite{Gluck:1998xa}
was
proposed. The main modifications are based on the  application of the
introduced new scaling variable $\xi_w$ (see Ref.\cite{Bodek}).
These  modified GRV98 LO PDFs can be used to model electron, muon
and neutrino inelastic scattering cross-sections at both very low
and high energies.

Among the most recent  problems discussed  at the  WG3 meeting, there
was the NuTeV anomaly,
described in the  detailed talk of Bernstein \cite{Bernstein}.
In this talk, the extracted $\sin^2\theta_W$ from NuTeV data
for the ratios of neutral current to charged current DIS cross-sections
\begin{equation}
R^{\nu}= \frac{\sigma(\nu N\rightarrow \nu X)}
{\sigma(\nu N\rightarrow l^-X)}~~{\rm and}~ R^{\overline{\nu}} =
\frac{\sigma(\overline{\nu} N\rightarrow \overline{\nu} X)}
{\sigma(\overline{\nu} N\rightarrow l^+X)}
\label{Ratio}
\end{equation}
is reported.
The NuTeV collaboration obtained the ratios $R^{\nu}=0.3916\pm 0.0007$
and $R^{\overline{\nu}}=0.4050\pm 0.0016$ from their data.
Using a LO description for the cross-sections with the LO PDFs,
they obtained the following value
for $\sin^2\theta_W$ \cite{Zeller:2001hh}:
\vfill\eject
\begin{eqnarray}
\! \! \! \! \! \! \! \! \! \! \! \! \! \! \! \! \! \! \! \! \! \! \! \!
\! \! \! \! \! \!
{\rm sin}^2\theta_W^{on-shell}&=&0.2277\pm 0.0013~
       ({\rm stat})\pm 0.009~(\rm{syst})
\nonumber \\
&-&0.00022 \bigg(\frac{M_t^2-(175~{\rm GeV})^2}{(50~{\rm GeV}^2)}\bigg)
\label{NuTeV}
+0.00032~{\rm ln}\bigg(\frac{M_{Higgs}}{150~{\rm GeV}}\bigg)~.
\end{eqnarray}
This value has a 3$\sigma$ deviation from the one obtained from the
fit to other electroweak measurements by LEPEWWG:
${\rm sin}^2\theta_W^{on-shell}=0.2227\pm 0.0037$.
The origin of the new NuTeV deviation from the standard LEPEWWG one
was tried to be understood in the talk of Davidson \cite{Davidson}, which
was based on the detailed work of Ref. \cite{Davidson:2001ji}.
Unfortunately no convincing  enough arguments in favor of
the explanation of the origin of the existing discrepancy
(see, however, Ref. \cite{Loinaz})
including the consideration  of definite  effects from physics beyond
the standard model, were found.
In this situation, it is worth while relying on
publication of NuTeV data for cross-sections and SFs
to allow the interested community to try to clarify the situation
with the help of independent analyses of these new experimental
data, which should include all  types
of   QCD effects, including those  related
to NLO perturbative QCD corrections.

To conclude, the work of WG3 at the NuFact'02 workshop turned out
to be valuable for various aspects of non-oscillation neutrino physics.
We hope that the outcome of these works will be used
for the continuation of the planning of new experiments
at the neutrino factories.

\vspace{0.4cm}
\noindent
{\bf Acknowledgments}

The conveners express their thanks to all the speakers of WG3 for
their interesting contributions. We are grateful to our co-convener
K. McFarland for his contributions to the WG3 organization.
SK thanks Y. Kuno for financial support for his participating
in this workshop. ALK was supported by RFBR Grants N 00-02-17432
and N 02-01-00601. He is also grateful to the organizers of this
workshop for the support of his stay in London and to
the Theory Division of CERN for hospitality during the work
on part of this summary.


\end{document}